# Temperature Dependent Product Yields for the Spin Forbidden Singlet Channel of the C($^3$P) + C$_2$H$_2$ Reaction

Kevin M. Hickson,[1,2]* Jean-Christophe Loison[1,2] and Valentine Wakelam[3,4]

[1]Univ. Bordeaux, ISM, UMR 5255, F-33400 Talence, France.
[2]CNRS, ISM, UMR 5255, F-33400 Talence, France.
[3]Univ. Bordeaux, LAB, UMR 5804, F-33270 Floirac, France.
[4]CNRS, LAB, UMR 5804, F-33270 Floirac, France.

The atomic hydrogen formation channels of the C + C$_2$H$_2$ reaction have been investigated using a continuous supersonic flow reactor over the 52–296 K temperature range. H-atoms were detected directly at 121.567 nm by vacuum ultraviolet laser induced fluorescence. Absolute H-atom yields were determined by comparison with the H-atom signal generated by the C + C$_2$H$_4$ reaction. The product yields agree with earlier crossed beam experiments employing universal detection methods. Incorporating these branching ratios in a gas-grain model of dense interstellar clouds increases the c-C$_3$H abundance. This reaction is a minor source of C3-containing molecules in the present simulations.

## 1. Introduction

Ground state atomic carbon, C($^3$P), (hereafter denoted C) is an important gas-phase species in both high temperature and low temperature environments [1]. In cold dense interstellar clouds where temperatures as low as 10 K are prevalent, C-atoms are thought to play a crucial role in the chemical complexification through barrierless reactions with a wide range of species [2]. Given its propensity to react with unsaturated hydrocarbons through addition to double or triple carbon bonds followed by hydrogen elimination [1], such processes are considered to be important mechanisms for the synthesis of long unsaturated carbon chain molecules in interstellar space. The reaction between C-atoms and acetylene, C$_2$H$_2$, is one of the most well studied gas-phase reactions of atomic carbon with three possible product channels at low temperature and pressure.

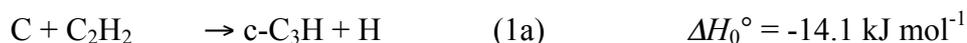
C + C$_2$H$_2$ → c-C$_3$H + H   (1a)   $\Delta H_0° = -14.1$ kJ mol$^{-1}$

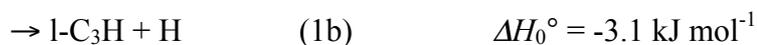
→ l-C$_3$H + H   (1b)   $\Delta H_0° = -3.1$ kJ mol$^{-1}$

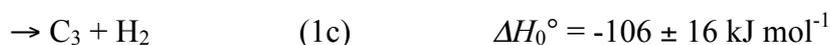
→ C$_3$ + H$_2$   (1c)   $\Delta H_0° = -106 \pm 16$ kJ mol$^{-1}$

A wide range of experimental studies of the kinetics [3,4] and dynamics [5-8] of reaction (1) exist, with numerous theoretical investigations attempting to explain the various experimental observations [9-14]. Rate constants for reaction (1) have been measured over a wide temperature range [3,4,15], proving that this process occurs over a barrierless potential energy surface for at least one of the three thermodynamically accessible pathways. One of the most interesting aspects of this reaction is the observation by several groups of large branching ratios for channel (1c) [6-8]; a process which can only take place by intersystem crossing between triplet and singlet potential energy surfaces of the intermediate $C_3H_2$ molecule. Leonori et al. [6] used the crossed molecular beam (CMB) scattering technique coupled with soft electron ionization mass spectrometry to obtain angular and translational energy distributions for products at masses m/z = 36 ($C_3$) and 37 (l-/c-$C_3H$) over a range of collision energies from 3.6 to 49.1 kJ mol$^{-1}$. Using these results, these authors were able to derive product yields for the $C_3$ + $H_2$ channel, (1c)/(1), and the one between the two H-atom production channels, (1a)/(1a)+(1b). They determined that channel (1c) represents a significant fraction of the total at low collision energy ((1c)/(1) = 0.5 ± 0.1 at 3.6 kJ mol$^{-1}$), which decreased with increasing collision energy. Similarly, the ratio (1a)/(1a)+(1b) was also found to decrease with increasing collision energy. The most recent experimental investigation of this reaction by Costes et al. [7], using a pulsed CMB method coupled with resonance enhanced multiphoton ionization to detect H-atoms, also determined the ratios (1a)/(1a)+(1b) over the range 0.44 - 4.5 kJ mol$^{-1}$ and (1c)/(1) at energies of 0.8 and 4.8 kJ mol$^{-1}$ (corresponding to the mean energies of thermal distributions at 64 and 385 K respectively). These authors derived values for (1a)/(1a)+(1b) which were mostly consistent with those determined by Leonori et al. [6]. To obtain the branching fractions for $C_3$ production (1c)/(1), Costes et al. [7] crossed a beam containing an equimolar mixture of $C_2H_2$ and $C_2H_4$ with a beam containing C-atoms, effectively using the C + $C_2H_4$ reaction as a reference. As the translational energies of $C_2H_2$ and $C_2H_4$ were essentially the same (similar reduced masses and identical relative velocities) giving rise to the same collision energy, they were able to deconvolute the respective H-atom signals in the resulting Doppler-Fizeau spectra (see Figure 7 of Costes et al. [7]) given the much lower exoergicity for H-atoms produced by the C + $C_2H_2$ reaction. They derived values of 0.82 and 0.87 for the ratio (1c)/(1) at 0.8 and 4.8 kJ mol$^{-1}$, considerably larger than the ones obtained by previous work in the same energy range [6,8], or at equivalent temperatures [15]. Moreover, the ratio (1c)/(1) was seen to increase with increasing collision energy; an observation which is seemingly at odds with theoretical considerations based on the argument that intersystem crossing is promoted by an increased lifetime of the $C_3H_2$ intermediate.

As the experiments of Costes et al. [7] are the only ones to have been performed in an energy range relevant to interstellar clouds at the present time, the product yields derived in their study have been universally adopted by astrochemical databases [16] as the preferred values, making reaction (1) an important source of $C_3$ molecules in astrochemical models [2]. As $C_3$ is thought to be unreactive with both atomic nitrogen [17] and oxygen [18,19] which are predicted to be present at high abundances in dense interstellar clouds, such high yields could inhibit complex molecule formation in current simulations. Indeed, both $l/c$-$C_3H$ produced by channels (1a) and (1b) are currently considered to react rapidly with both of these atomic species [17].

To resolve the current discrepancies between these recent studies, we investigated reaction (1) using a continuous supersonic flow apparatus to derive product yields for channel (1c) over the 52 - 296 K range. The experimental methods used are described in section (2). The results are presented and discussed in section (3) while the astrophysical implications of these results are outlined through modeling studies in section (4). Our conclusions are given in section (5).

## 2. Experimental Methods

The H-atom production channels of the $C + C_2H_2$ reaction were investigated using a small continuous supersonic flow reactor [20-23]. Three Laval nozzles were used during the course of this study, allowing flow temperatures in the range 52-177 K to be accessed. The measured and calculated values for these nozzles are given in Table 1.

**Table 1** Characteristics of the supersonic flows

| Mach number | $1.8 \pm 0.02$[a] | $3.0 \pm 0.02$ | $3.9 \pm 0.1$ |
|---|---|---|---|
| Carrier gas | $N_2$ | $N_2$ | Ar (8% $N_2$) |
| Density ($\cdot 10^{16}$ cm$^{-3}$) | 9.4 | 10.3 | 25.9 |
| Impact pressure (Torr) | 8.2 | 13.4 | 29.6 |
| Stagnation pressure (Torr) | 10.3 | 39.7 | 113 |
| Temperature (K) | $177 \pm 2$ | $106 \pm 1$ | $52 \pm 1$ |
| Mean flow velocity (ms$^{-1}$) | $496 \pm 4$ | $626 \pm 2$ | $505 \pm 1$ |

In addition, experiments were conducted at 296 K by operating without a Laval nozzle, using $N_2$ as the carrier gas at a total pressure of approximately 5 Torr.

$C(^3P)$ atoms were produced by the dissociation of precursor $CBr_4$ molecules (a multiphoton process) using the 10 Hz pulsed laser photolysis method with approximately 20 mJ at 266 nm. This process is also known to produce a small fraction of excited state $C(^1D)$ atoms at the level of 10-15 % with respect to $C(^3P)$ [20].

The photolysis laser was aligned along the reactor axis, creating carbon atoms within the supersonic flow. $CBr_4$ molecules were carried into the reactor by a small $N_2$ or Ar flow which was passed into a flask containing solid $CBr_4$ at a known pressure, upstream of the nozzle reservoir. From its saturated vapor pressure at room temperature, we estimate that the maximum $CBr_4$ concentration used in these experiments was $5 \times 10^{12}$ molecule $cm^{-3}$.

Product $H(^2S)$ atoms were detected by vacuum ultraviolet laser induced fluorescence (VUV LIF) using the 1s $^2S \rightarrow$ 2p $^2P^0$ Lyman-α transition at 121.567 nm and a VUV photomultiplier tube (PMT). To produce light at and around this wavelength, a pulsed dye laser operating at 729.4 nm was frequency doubled to produce tunable light around 364.7 nm. This beam was focused into a cell positioned perpendicularly to the cold flow and the PMT containing 210 Torr of krypton with 540 Torr of argon added for phase matching. The tunable VUV light produced by third harmonic generation was collimated using a $MgF_2$ lens before crossing the supersonic flow.

In order to determine absolute H-atom product yields, H-atom VUV LIF signal intensities from the $C(^3P) + C_2H_2$ reaction were compared to those generated by the reference $C(^3P) + C_2H_4$ reaction; a process with a known H-atom yield of $0.92 \pm 0.04$ at 300 K [15] which we consider to be constant over our experimental temperature range. To ensure that reactions of $C(^1D)$ atoms with $C_2H_2$ and $C_2H_4$ did not produce supplementary H-atoms (thereby distorting the measured yields), Laval nozzles employing $N_2$ as the carrier gas were used. An Ar based nozzle was used to study reaction (1) at the lowest temperature (52 K), however in this case a large concentration of $N_2$ ($2.1 \times 10^{16}$ molecule $cm^{-3}$) was added to the flow. A recent study of the $C(^1D) + N_2 \rightarrow C(^3P) + N_2$ quenching reaction has shown that this process rapidly removes $C(^1D)$ atoms, becoming more efficient at low temperature [24] reaching a value around 1.5 ×

$10^{-11}$ cm$^3$ molecule$^{-1}$ s$^{-1}$ at 50 K. As a result, all of the C($^1$D) atoms are removed under our experimental conditions within the first few microseconds.

VUV LIF signals were recorded as a function of delay time between the photolysis laser and the probe laser. To ensure that diffusional losses did not lead to large errors in the estimation of the H-atom yields, only profiles with similar time constants were used to extract relative intensities. As a result, excess reagent concentrations in the range (5.3 – 14.6) × 10$^{13}$ molecule cm$^{-3}$ (C$_2$H$_2$ or C$_2$H$_4$) were chosen to obtain similar first-order-production rates for both reactions. The rate constants for both of these reactions are virtually identical, with values of (3.0 – 3.6) × 10$^{-10}$ cm$^3$ molecule$^{-1}$ s$^{-1}$ over the 50-300 K temperature range [4]. For a given coreagent concentration, 30 datapoints were acquired at each time step with at least 50 time intervals for each formation curves. In addition, several points were recorded at negative time delays to set the baseline level.

Gases were flowed directly from cylinders without purification prior to use (Ar Linde 99.999%, N$_2$ Air Liquide 99.999%, C$_2$H$_4$ Linde 99.9%, C$_2$H$_2$ Messer 99.6%). As the C$_2$H$_2$ used was supplied dissolved in acetone for stability, we also performed a secondary mass spectrometric verification of its purity. This analysis indicated that the acetone content of the C$_2$H$_2$ gas was at most 1.4 %. Digital mass flow controllers were used to control the carrier gas flows. These controllers were calibrated prior to usage for the specific gas used.

## 3. Results and Discussion

Typical H-atom formation curves recorded sequentially for the C + C$_2$H$_2$ and C + C$_2$H$_4$ reactions are shown in Figure 1 for experiments performed at 52 K (panel A) and 296 K (panel B) respectively.

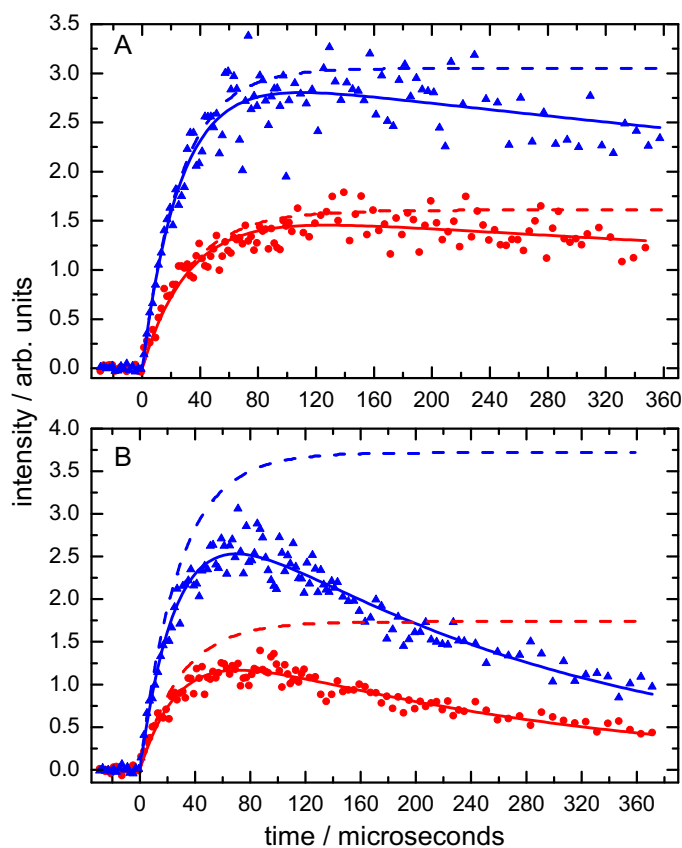

**Figure 1.** H-atom VUV LIF emission profiles recorded at (A) 52 K and (B) 296 K. (Blue solid triangles) H-atom signal from the C + $C_2H_4$ reaction with $[C_2H_4] = 1.3 \times 10^{14}$ molecule $cm^{-3}$ at 52 K and $1.5 \times 10^{14}$ molecule $cm^{-3}$ at 296 K; (blue solid line) biexponential fit to the C + $C_2H_4$ datapoints; (blue dashed line) theoretical H-atom yield from the C + $C_2H_4$ reaction in the absence of competing H-atom losses. (Red solid circles) H-atom signal from the C + $C_2H_2$ reaction with $[C_2H_2] = 1.3 \times 10^{14}$ molecule $cm^{-3}$ at 52 K and $[C_2H_2] = 1.5 \times 10^{14}$ molecule $cm^{-3}$ at 296 K; (red solid line) biexponential fit to the C + $C_2H_2$ datapoints; (red dashed line) theoretical H-atom yield from the C + $C_2H_2$ reaction in the absence of competing H-atom losses.

Several pairs of decays similar to the ones shown were recorded at each temperature, to minimize potential experimental errors. Additionally, the order in which the traces were acquired was alternated to reduce errors arising from possible changes in the fluorescence intensities over time. The curves are described by two component parts; an initial rapid rise, due to H-atom formation followed by slow H-atom loss, mostly by diffusion from the probe volume. A biexponential function was used to describe the evolution of the H-atom signal intensity ($I_H$) with an exponential loss term $k_{L(H)}$ to describe secondary H-atom loss.

$$I_H = A\{\exp(-k_{L(H)}t) - \exp(-k_{1st}t)\} \qquad (1)$$

The first-order formation rate, $k_{1st} = k_{C+X}[X] + k_{L(C)}$ where X represents either $C_2H_2$ or $C_2H_4$. The H-atom formation rate also depends on the first-order losses of C-atoms (such as through diffusion and secondary reactions) $k_{L(C)}$, making it important to use similar first-order-production rates for both reactions, as some carbon atoms might be lost without reacting with either of the coreagents. The parameter $A$ in equation (1) represents the theoretical amplitude of the H-atom VUV LIF signal without secondary H-atom losses. In Figure 1, we also show the the H-atom signal obtained when $k_{L(H)}$ is set to zero in equation (1) (ie the theoretical H-atom yield). The ratio of the $A$ parameter values derived from the fits thus represents the relative H-atom yields between the target and reference reactions. Two factors must be considered to obtain accurate absolute H-atom yields for the $C + C_2H_2$ reaction. Firstly, as the H-atom yield of the $C + C_2H_4$ reaction (with $C_3H_3$ as the coproduct) is not 100 % (0.92 ± 0.04 at 300 K [15]), a correction factor must be applied for the reference reaction. The $C + C_2H_4$ reaction leads to 3 surfaces in the entrance valley. At the MRCI+Q level, with the CASSCF geometry optimized for non-relaxed $C_2H_4$ (the geometry of the isolated molecule), only one of the three surfaces is attractive, leading to cyclic-$C_3H_4$ (214 kJ mol$^{-1}$ below the reagent level). The evolution of this cyclic species has already been described by Le et al. [25], leading mainly to $C_3H_3$ + H through various steps involving transition states which are at least 171 kJ mol$^{-1}$ below the reagents. As the initial cyclic-$C_3H_4$ intermediate is very low with respect to the entrance channel and the evolution of this cyclic species involves only very low submerged barriers, the H-atom branching ratio close to 1 is not expected to display a significant variation with temperature. Secondly, we need to consider possible absorption of the VUV excitation source and fluorescence emission by residual gases ($C_2H_2$ and $C_2H_4$ in particular) in the reactor. Secondary absorption was found to be as high as 8 % at 296 K for experiments conducted with $C_2H_2$, with a corresponding 5 % correction for $C_2H_4$ experiments. Appropriate corrections were applied to the measured intensities although below room temperature, the correction was less than 2 % for all experiments. The absolute temperature dependent H-atom branching ratios obtained for the $C + C_2H_2$ reaction are listed in Table 2. These yields are derived from the mean ratio of at least eight pairs of $A$ factors at each temperature.

**Table 2** Temperature dependent H-atom yields for the $C(^3P) + C_2H_2$ reaction

| T / K | Number of experiments | Individual H-atom | Mean H-atom yield |
| --- | --- | --- | --- |

|     |    | yields |     |
| --- | -- | ------ | --- |
| 296 | 9  | 0.43, 0.41, 0.39, 0.42, 0.40, 0.45, 0.41, 0.44, 0.40 | 0.42 ± 0.02[a] |
| 177 | 8  | 0.45, 0.48, 0.44, 0.47, 0.48, 0.45, 0.43, 0.46 | 0.46 ± 0.02 |
| 106 | 13 | 0.45, 0.48, 0.52, 0.40, 0.43, 0.49, 0.49, 0.39, 0.50, 0.48, 0.44, 0.44, 0.46 | 0.46 ± 0.03 |
| 52  | 8  | 0.50, 0.49, 0.46, 0.46, 0.50, 0.49, 0.46, 0.44 | 0.47 ± 0.03 |

[a]The error bars reflect the statistical uncertainties at the 95 % confidence level including the uncertainties of the H-atom yield of the reference C + $C_2H_4$ reaction.

The H-atom yields listed in Table 2 represent the branching fraction (1a)+(1b)/(1). The ratio for the $C_3$ + $H_2$ channel, (1c)/(1) is thus 1 - (1a)+(1b)/(1). These values are presented in Figure 2 alongside earlier work. Here, the results of CMB experiments at well-defined collision energies have been converted to temperature by assuming that these values are equivalent to the mean energies of thermal distributions.

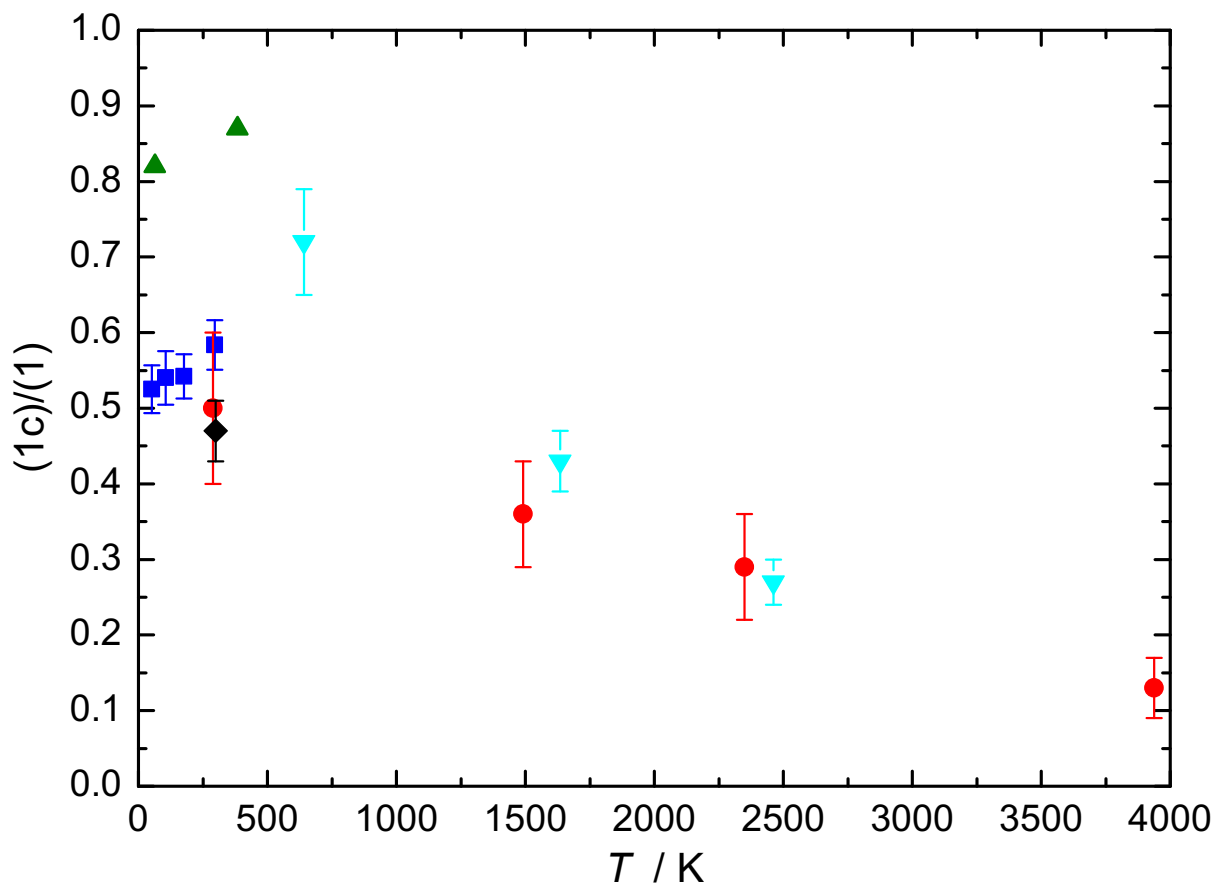

**Figure 2** Temperature dependent experimental branching ratios for channel (1c) of the C + $C_2H_2$ reaction. (Light blue solid triangles) Gu et al. [8]; (Red solid circles) Leonori et al. [6]; (Black solid diamond) Bergeat and Loison [15]; (Green solid triangles) Costes et al. [7]; (Blue solid squares) this work.

The room temperature branching ratio for channel (1c) of 0.58 ± 0.03 determined in this study agrees well with the earlier result of Bergeat and Loison of 0.47 ± 0.04 at 300 K [15] and the result of Leonori et al. [6] of 0.50 ± 0.1 at a collision energy of 3.6 kJ mol$^{-1}$ (≈ 290 K). The value derived by Costes et al. [7] at a collision energy of 4.8 kJ mol$^{-1}$ (≈ 385 K) is 50 % higher, indicating possible errors in the deconvolution of the H-atom signals from the target C + $C_2H_2$ and reference C + $C_2H_4$ reactions in their spectral analysis. Our results show that the branching ratio for channel (1c) decreases slightly at lower temperature, reaching a value of 0.53 ± 0.03 at 52 K. While the overall magnitude of the branching ratios (1c)/(1) derived by Costes et al. [7] seems to be erroneous, the energy (temperature) dependence of these results is nonetheless similar to the one determined in this work, indicating that the $C_3$ + $H_2$ product channel becomes less favourable as the temperature falls over the 300 – 50 K range. In contrast, the earlier studies of Leonori et al. [6] and Gu et al. [8] both found that the branching ratio for channel (1c) increases to low energy. While these results seem to contradict the present findings at first glance, these datasets might still be consistent. Indeed, the branching ratios measured at high collision energies by both Leonori et al. [6] and Gu et al. [8] are very similar, whereas the value measured by Gu et al. [8] at 8.0 kJ mol$^{-1}$ is somewhat larger than what might be expected by comparison with the low energy data of Leonori et al. [6], the room temperature value of Bergeat and Loison [15] and those determined here. Nevertheless, this value could simply be indicative of a turnover in the branching fraction for this process in the 4 – 15 kJ mol$^{-1}$ range of collision energies. Further CMB measurements in this intermediate range of collision energies (or kinetic measurements of the H-atom yields in the 300 – 1200 K temperature range) should allow this hypothesis to be verified.

Other possible explanations for the discrepancy between the temperature dependences might be attributed to differences in the respective spin-orbit populations of C($^3P_{0,1,2}$) atoms between the present bulk type of experiments and those produced in CMB apparatuses. While a thermal population is always maintained between the $J$ = 0, 1, 2 spin-orbit levels (with energies of 0, 16.4 and 43.4 cm$^{-1}$ respectively) in the present study due to rapid equilibration through collisions with the carrier gas molecules, the same cannot be said of the distribution of populations in molecular beam experiments. Indeed, due to the typical low molecular beam densities, the nascent population distribution of C-atoms is expected to be preserved in the beam crossing region. Takahashi and Yamashita [10] show that of the three triplet electronic states correlating with C($^3P$) + $^1C_2H_2$ reagents, only one presents no barrier. Considering the fine structure of atomic carbon, only the $^3P_0$ state and two levels of the $^3P_1$ states of the C-atom lead to reaction. As a result, the significantly different spin-orbit level populations generated by these experimental methods could induce notable differences in the reactivity, particularly through nonadiabatic couplings which might become accessible at higher collision energies.

Another similar issue relates to the coreagent $C_2H_2$ molecules in CMB experiments. These species are likely to be strongly rotationally cooled by the initial supersonic expansion, leading to a significantly different rotational temperature than the one found in kinetic type experiments at equivalent temperatures. As before, a differing reactivity for the initial rotational states of the coreagent molecules could lead to large differences between results obtained by these methods.

**4. Astrophysical Implications**

In order to test the effect of the revised branching ratios on simulations of interstellar regions, we used the chemical model Nautilus [26,27]. Using a set of initial physical and chemical parameters, this program simulates the chemical abundances as a function of time, considering gas-phase reactions, species exchange between the gas-phase and interstellar grain surfaces and reactions on the grain surfaces themselves. The numerical model and chemical network used for these simulations are the same as in Hickson et al. [18]. We start with an initial chemical composition representative of a diffuse cloud (i.e. all species are in atomic or ionic form except for hydrogen, which is molecular), and use typical cold core physical conditions [18]. In addition to this standard model, we have run a model in which we changed the rate constants for the individual channels of the $C + C_2H_2$ reaction to reflect the values obtained in this work; $k_{1a}(T) = 1.2\text{e-}10 \times (T/298)^{-0.18}$ cm$^3$ molecule$^{-1}$ s$^{-1}$, $k_{1b}(T) = 1.2\text{e-}11 \times (T/298)^{1.08}$ cm$^3$ molecule$^{-1}$ s$^{-1}$ and $k_{1c}(T) = 1.7\text{e-}10 \times (T/298)^{-0.08}$ cm$^3$ molecule$^{-1}$ s$^{-1}$. A third model was also run with the $C + C_2H_2$ reaction switched off to test its overall importance.

Figure 3 shows the abundances of $C_2H_2$, $C_3$, c-$C_3H$ and l-$C_3H$ computed by the three models in the gas-phase as a function of time.

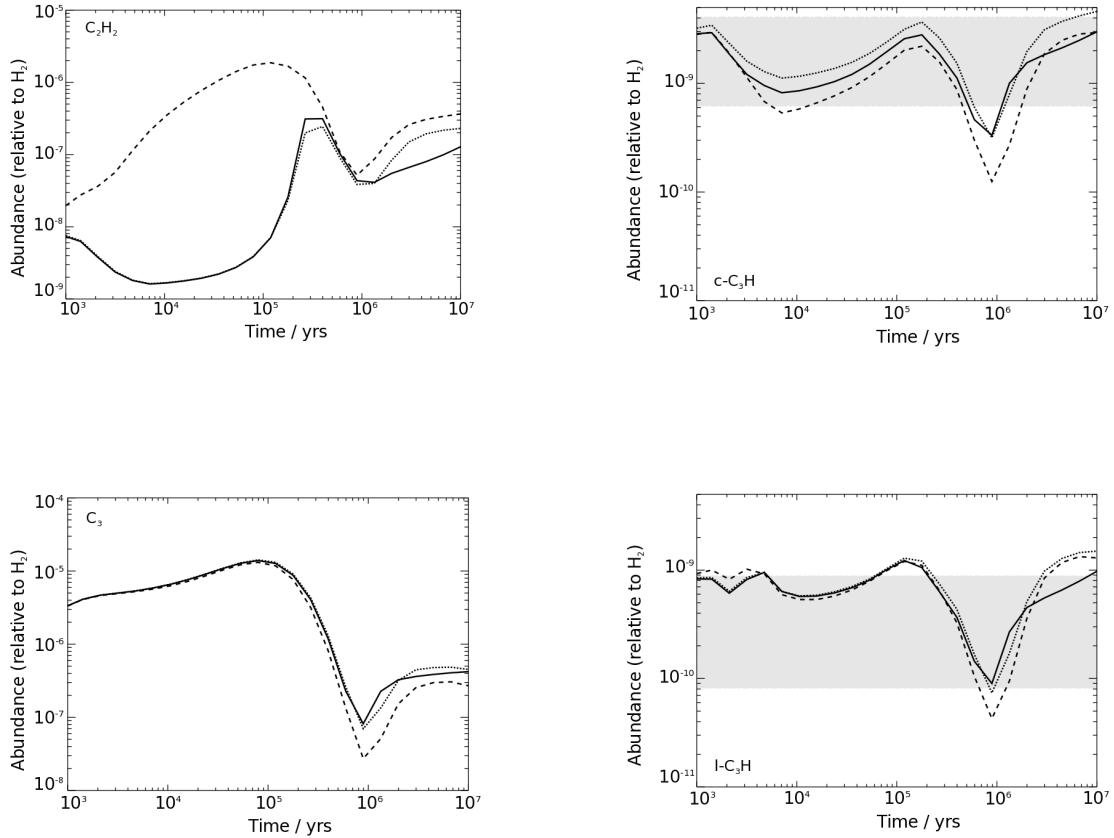

**Figure 3** Abundances of $C_2H_2$, $C_3$, c-$C_3H$ and l-$C_3H$ computed by the three models in the gas-phase as a function of time. (Solid line) branching ratios for the $C + C_2H_2$ reaction based on the study by Costes et al. [7]; (dotted line) new branching ratios for the $C + C_2H_2$ reaction; (dashed line) with the $C + C_2H_2$ reaction switched off. The horizontal grey zones represent the various observations for TMC-1 (CP). [28]: c-$C_3H$/$H_2$ = 1 × $10^{-9}$ and l-$C_3H$/$H_2$ = 8 × $10^{-11}$, [29]: c-$C_3H$/$H_2$ = 7 × $10^{-10}$, [30] l-$C_3H$/$H_2$ = 9 × $10^{-10}$, [31]: c-$C_3H$/$H_2$ = 3 × $10^{-9}$ and l-$C_3H$/$H_2$ = 6 × $10^{-10}$.

The main effect of the new branching ratios is an increase of the c-$C_3H$ abundance whilst the l-$C_3H$ abundance shows little variation, indicating that there are other more important sources of this radical. The main effect in the suppression of the $C + C_2H_2$ reaction, is a massive enhancement of the $C_2H_2$ abundance (as the $C + C_2H_2$ reaction is the main $C_2H_2$ loss process). The change in the branching ratios presented in this work has only a small influence on the predicted abundances. This observation is mainly due to the particular behavior of the $C_3$ molecule. In our network, there are various efficient pathways producing $C_3$ with very few efficient destruction mechanisms. At early times, $C_3$ is produced by a sequence of reactions beginning with the $C^+ + C_2H \rightarrow C_3^+ + H$ reaction followed by $C_3^+ + H_2 \rightarrow C_3H^+ + H$ and dissociative electron recombination of $C_3H^+$ to give neutral $C_3$ at high abundance levels ($10^{-5}$ with respect to [$H_2$] at $10^5$ years). By comparison, $C_3$ production through the $C + C_2H_2$ reaction is insignificant at these times. At later times (> 4 × $10^5$ years), the $C + C_2H_2$ reaction becomes the major source of $C_3$ although its abundance is two orders of magnitude less than the peak value. $C_3$ has a low reactivity with abundant species in molecular clouds such as O, H, N, CO, $CH_4$ and $C_2H_2$ [19,32-37]. $C_3$ reacts without a barrier with C-atoms leading to $C_4$ through radiative association [38] with a small flux. As $C_4$ leads mainly back to $C_3$ through reactions with C, O and N atoms [38], the only efficient reactions for $C_3$ loss are with $H_3^+$, $HCO^+$ and $HCNH^+$ and through depletion onto grains. Then the high abundance of $C_3$ leads to efficient $C_3H^+$ production through proton transfer and then to efficient c-/l-$C_3H_2^+$ and c-/l-$C_3H_3^+$ production through the $C_3H^+ + H_2$ reaction [39-41]. Dissociative electronic recombination reactions of c-/l-$C_3H_2^+$ and c-/l-$C_3H_3^+$ drive the production of c-/l-$C_3H$ and c-/l-$C_3H_2$ as these processes involve substantially higher fluxes than the $C + C_2H_2$ reaction. In a similar manner to the simulated $C_3H_6$ abundance [18], the predicted presence of a barrier for the $O + C_3$ reaction [19] is critical for c-/l-$C_3H$ and c-/l-$C_3H_2$ abundances.

**5. Conclusions**

Branching ratios for the H-atom production channels of the $C + C_2H_2$ reaction have been measured using a supersonic flow apparatus over the 52 – 296 K range. These values allow us to determine branching fractions for the spin forbidden singlet channel which display a weak temperature dependence, decreasing slightly as the temperature falls. The overall magnitude of this branching ratio is in excellent agreement with earlier kinetic measurements at 300 K and crossed molecular beam measurements employing universal mass spectrometric detection. The most recent crossed molecular beam measurements of the branching ratio for the spin forbidden channel using spectroscopic detection of H-atoms are inconsistent with the present and previous work and are therefore likely to be erroneous.

A modeling study of the effect of the $C + C_2H_2$ reaction on dense cloud chemistry indicates that this process is only a minor source of the radicals $C_3$ and l-/c-$C_3H$ at early times, despite being the major sink for interstellar $C_2H_2$. The new branching ratios lead to a noticeable increase in the c-$C_3H$ abundance while the abundances of l-$C_3H$ and $C_3$ are relatively unchanged.

**Acknowledgements**

The authors acknowledge support from the French INSU/CNRS Programs 'Physique et Chimie du Milieu Interstellaire' (PCMI) and 'Programme National de Planétologie' (PNP). V.W.'s research is funded by the ERC Starting Grant (3DICE, grant agreement 336474).

*Corresponding author at: Université de Bordeaux, ISM, UMR 5255, F-33400 Talence, France

E-mail address: kevin.hickson@u-bordeaux.fr (K.M. Hickson)